\documentclass[aps,twocolumn,pra,superscriptaddress]{revtex4-2}
\usepackage{amsfonts}
\usepackage[final,hiresbb]{graphicx}
\usepackage{amsmath}
\usepackage{amssymb}
\usepackage{amstext}
\usepackage{color}
\usepackage{braket}

\usepackage{subfigure}
\usepackage{appendix}
\usepackage{bbm}
\usepackage{mathbbol}

\definecolor{Green}{RGB}{0,204,102}
\definecolor{Purple}{RGB}{102,0,255}
\definecolor{Blue}{RGB}{51,153,255}
\definecolor{Red}{RGB}{255,010,010}

\newcommand{\phii}{\phi_i}
\newcommand{\qi}{\theta_i}
\newcommand{\el}{\mathcal{l}}

\newcommand{\pgE}{\Phi^{\rm geom}_{\rm\Psi}}
\newcommand{\ptE}{\Phi^{\rm tot}_{\rm\Psi}}
\newcommand{\pdE}{\Phi^{\rm dyn}_{\rm\Psi}}
\newcommand{\GPhi}{{\rm G_{\rm \scriptstyle \Phi}}}
\newcommand{\Gproj}{{\rm G_{\rm \scriptstyle \mathbb{P}}}}
\newcommand{\GPhisolo}{\rm G_{\rm \scriptstyle \Phi}^{\rm solo}}
\newcommand{\Gprojsolo}{\rm G_{\rm \scriptstyle \mathbb{P}}^{\rm solo}}

\newcommand{\pgA}{\Phi^{\rm geom}_A}
\newcommand{\pgB}{\Phi^{\rm geom}_B}

\newcommand{\Paa}{\mathbb{P}_{\!\! {\scriptscriptstyle A\!A}}}
\newcommand{\Pab}{\mathbb{P}_{\!\! {\scriptscriptstyle A\!B}}}
\newcommand{\Pba}{\mathbb{P}_{\!\! {\scriptscriptstyle B\!A}}}
\newcommand{\Pbb}{\mathbb{P}_{\!\! {\scriptscriptstyle B\!B}}}
\newcommand{\PPsi}{\mathbb{P}_{\! {\rm \Psi}}}
\newcommand{\PS}{\mathbb{P}_{\! {\rm S}}}

\newcommand{\PPsisolo}{\mathbb{P}_{\! E, {\rm solo}}}
\newcommand{\PSsolo}{\mathbb{P}_{\! S, {\rm solo}}}

\newcommand{\ketp}{\ket{\scriptscriptstyle  +}}
\newcommand{\ketm}{\ket{\scriptscriptstyle -}}

\newcommand{\PhitotA}{\Phi_{\!\! {\scriptscriptstyle A}}^{\rm tot}}
\newcommand{\PhitotB}{\Phi_{\!\! {\scriptscriptstyle B}}^{\rm tot}}

\newcommand{\PhidynA}{\Phi_{\!\! {\scriptscriptstyle A}}^{\rm dyn}}
\newcommand{\PhidynB}{\Phi_{\!\! {\scriptscriptstyle B}}^{\rm dyn}}
\newcommand{\PhidynAB}{\Phi_{\!\! {\scriptscriptstyle A\!B}}^{\rm dyn}}
\newcommand{\PhidynBA}{\Phi_{\!\! {\scriptscriptstyle B\!A}}^{\rm dyn}}

\newcommand{\PhigeomA}{\Phi_{\!\! {\scriptscriptstyle A}}^{\rm geom}}
\newcommand{\PhigeomB}{\Phi_{\!\! {\scriptscriptstyle B}}^{\rm geom}}

\newcommand{\Psigen}{\Psi_{\scriptscriptstyle {\rm 0}}}
\newcommand{\gammagen}{\gamma_{\scriptscriptstyle {\rm 0}}}
\newcommand{\Psigengauge}{\breve{\Psi}_{\scriptscriptstyle {\rm 0}}}

\newcommand{\Projgen}{\mathbb{P}_{\scriptscriptstyle{\rm 0}}}
\newcommand{\Projgengauge}{\breve{\mathbb{P}}_{\scriptscriptstyle{\rm 0}}}

\newcommand{\Phidyngen}{\Phi_{\!\! {\scriptscriptstyle {\rm 0}}}^{\rm dyn}}
\newcommand{\Phidyngengauge}{\breve{\Phi}_{\!\! {\scriptscriptstyle {\rm 0}}}^{\rm dyn}}

\usepackage{mathrsfs}
\usepackage[scr=esstix,cal=boondox]{mathalfa} 

\begin{document}

\title{The Influence of Quantum Correlation \\on the Holonomy of Spatially-Structured Bi-Photons}
	
\author{Mark T. Lusk}
	\email{mlusk@mines.edu}
	\affiliation{Department of Physics, Colorado School of Mines, Golden, CO 80401, USA}

	\begin{abstract}
The manifestation of entanglement within geometric phase is elucidated for spatially-structured bi-photons. Entanglement parameters are shown to influence holonomy in two distinct ways: through statistical superpositions of separable states; and via quantum correlation. These are entwined within geometric phase, motivating the construction of a projective, gauge-invariant measure that allows the manifestation of quantum correlation to be pinpointed and explained.  An optical circuit consisting of a pair of misoriented  mode converters gives a practical demonstration. This is facilitated by a novel pump engineering method which produces photon pairs with tunable entanglement.
\end{abstract}
	\maketitle

	\section{Introduction}
The interplay of quantum correlation and holonomy is fundamentally intriguing and technologically promising. Quantum optics offers an ideal setting to tease out physical understanding in this arena due to exquisite field control and relatively low  interactions. A two-stage methodology is called for: a means for tailoring the entanglement that gives rise to the geometric effects of interest, and a theoretical framework to pinpoint the way in which such entanglement is embodied in holonomy. Both of these are taken up in the present work using spatially structured bi-photons.

The ability to control the degree of multi-particle entanglement is an essential tool in studies of the foundations of quantum mechanics, including violations of local-realism\cite{Bell_1964}, teleportation\cite{Kok_2000}, and entanglement swapping \cite{Brukner_2005, Ma_2012}. Tunable entanglement also features prominently in current quantum sensing and computing technologies, because environmental interactions rapidly degrade entanglement. States engineered with partial entanglement are used to assess environmental effects\cite{vanExter_2006, Gomez_2018}, but the ability to work with such states directly is also being considered because they may be more robust to environmental degradation\cite{Lee_2000}. In settings governed by photon polarization, tunable entanglement can be produced using two birefringent crystal crystals with orthogonal optical axes\cite{White_1999,White_2001}. More recent work allows this to also be accomplished in an on-chip setting\cite{Maunz_2007, Setzpfandt_2016, Barral_2017}. 

The second cornerstone of the present work also has its roots in polarization studies, where holonomy was first studied theoretically by Pancharatnam\cite{Pancharatnam_1956}, followed by experimental measurements at the classical level\cite{Tomita_1986} and then with single-photon measurements\cite{Kwiat_1991}. This was subsequently generalized to tensor-valued, non-Abelian gauge fields\cite{Wilczek_1989, Pachos_2000}. Such geometric holonomies have subsequently been extended to non-adiabatic dynamics~\cite{Aharonov_1987} and open trajectories on the relevant parameter space~\cite{Samuel1988, Mukunda1993}. Here there is still closure, via geodesics, so the holonomy designation is still apt. They have been experimentally measured in both classical and quantum optics\cite{Simon_1988, Kwiat_1991, Chiao_1995, Galvez2003, Ericsson_2005, Franke_2002, Voitiv_2024}.  

Photonic measurements of geometric phase necessarily embraced entangled states to herald coincidence measurements, and this led naturally to a theoretical investigation of the influence of partial polarization entanglement on holonomy\cite{Hessmo_2000, Ericsson_2005}. The relation between the geometric phase of bipartite systems and their subsystems has also been considered\cite{Tong_2003}. These studies have shown that entanglement does, indeed, influence holonomy. This is potentially useful for quantum sensing and has immediate application to holonomic quantum computing, which seeks to harness the geometry of a parameter space to encode information with increased resilience to noise\cite{Sjoqvist_2000, Pachos_2000, Pinske_2023}. 

Structured light\cite{Andrews_2011} can also exhibit holonomy, and this has followed a development similar to that of polarization. Here the relevant parameter space is typically that of first-order, orbital angular momentum modes\cite{Padgett_1999} instead of the Poincar{\'e} sphere. The associated geometric phase holonomy has been thoroughly studied at both the classical\cite{vanenk1993, Galvez2003,Lusk_2022, Voitiv_2023} and quantum\cite{Mair_2001, Franke_2002, Voitiv_2024} levels. To explain the relationship between entanglement and holonomy in this setting, though, a methodology for constructing tunably entangled states is called for. This would facilitate an investigation of how holonomy is manifested when structured photons are entangled.

After briefly reviewing the geometric phase produced by single-photon orbital angular momentum (OAM) states, the analysis of a general class of entangled antipodal modes is used to show that the manifestation of entanglement in holonomy comes in two distinct forms. The first is via statistical mode weighting with an effect that could be achieved by a superposition of the dynamics of separated photons of each mode. The second manifestation, truer to the core idea of entanglement, is associated with the quantum correlation between modes. A gauge-invariant measure based on the difference of two geometric phases, a \emph{Geometric Phase of Entanglement} ($\GPhi$), simplifies the analysis by removing dynamic phase and path dependence from consideration. This lends insight but still includes both flavors of entanglement influence. This motivates the derivation of a gauge-invariant projective measure, the \emph{Geometric Projection of Entanglement} ($\Gproj$), that contains only the influence of quantum correlation. It has a particularly simple and useful interpretation, quantifying the projection of the initial state of one mode onto the final state of another---a signature feature of entanglement. 

Requisite states are produced through a novel methodology in which pump photons are engineered, prior to their encounter with a birefringent crystal, to produce tunably entangled, structured light. Specifically, pump photons are engineered as a linear combination Laguerre-Gaussian modes with well-defined orbital angular momenta (OAM), and subsequent Type-I Spontaneous Parametric Downconversion (SPDC) results in two-photon states that exhibit OAM mode entanglement. Tunably entangled states are then sent through a holonomy-producing optical circuit. The results identify the precise manner in which $\Gproj$ identifies the influence of entanglement on the manifestation of holonomy. The work is theoretical and includes analytical simulations, but the approach lends itself to straightforward experimental implementation.

A measure analogous to $\Gproj$ was very recently introduced in a general relativistic setting, where entangled polarization states were studied after transiting a closed three-space circuit around a rotating black hole\cite{Lusk_2024b}. The propagation of light was restricted to geodesics, so dynamic phase was not considered, entanglement was taken to be maximal, and only projective holonomies were investigated. 

\section{Review: Single-Photon Holonomy}

Structured light composed of linear combinations of Laguerre-Gaussian (LG) modes, $\ket{\el}$, may accumulate geometric phase as it passes through linear optical elements~\cite{vanenk1993,Padgett1999}.  Here $\el$ refers to the OAM parameter of a given mode, with radial parameter set to zero to give optical vortices.

The associated parameter space can be projected onto the first-order Sphere of Modes (SoM) (sometimes referred to as the modal or optical Poincar\'e Sphere)\cite{Padgett_1999}, so SU(2) curvature underlies holonomic measures. A convenient basis for the underlying states is a pair of LG modes with OAM parameters of $\el = \pm 1$. These are denoted $\{ \ketp, \ketm \}$. Paths on the SoM can then be described in terms of the evolution of scalar functions for polar and azimuthal angles, $\theta (s)$ and $\phi (s)$, respectively. This is an incomplete description of state evolution, remedied by generalizing the SoM to a fiber bundle in which a fiber extends out from each point on its surface. \emph{Horizontal lift}, $\chi(s)$, along the fibers tracks the change in total phase of the beam~\cite{Kobayashi_1996, Lusk_2022}. An evolving state, $\ket{A(s)}$, is therefore be represented as
\begin{equation}\label{stateA}
\ket{A} = e^{\imath \chi} \left(e^{-\imath \frac{\phi}{2}} \cos\left(\frac{\theta}{2}\right) \ketp + e^{\imath \frac{\phi}{2}}\sin\left(\frac{\theta}{2}\right)  \ketm\right) .
\end{equation}

A series of optical elements can be used to collectively trace out a path on the SoM. In terms of path parametrization, $s$, define a projection of initial state $\ket{A(s_i)}$ onto final state $\ket{A(s_f)}$ as
\begin{equation}\label{proj_single}
\Paa = \braket{A(s_i) | A(s_f)}.
\end{equation}
The total phase is then
\begin{equation}\label{phitot_single}
\Phi_{\rm tot,A} = \arg(\Paa).
\end{equation}
This is not gauge invariant, of course, but it will prove useful to derive a correction term which renders it so\cite{Mukunda_1993}. Introduce a gauge-transformed field,
\begin{equation}\label{Axform}
\ket{\breve A(s)} := e^{\imath \gamma(s)}\ket{A(s)},
\end{equation}
with $\gamma$ a smooth but otherwise arbitrary scalar function. Take the argument of the inner product of $\breve A$ at two positions to obtain
\begin{equation}\label{AR2}
\int_{s_i}^{s_f} ds \, \partial_s \gamma(s) = \arg\braket{\breve A(s_i) | \breve A(s_f)} - \arg\braket{A(s_i) | A(s_f)}.  
\end{equation}
Holding onto this relation for a moment, take derivative of Eq. \ref{Axform} to obtain
\begin{align}\label{AR3}
\int_{s_i}^{s_f} ds \, \partial_s \gamma(s) = &-\imath \int_{s_i}^{s_f} ds \, \braket{\breve A(s) | \partial_s\breve A(s)} \nonumber \\
& + \imath \int_{s_i}^{s_f} ds \, \braket{A(s) | \partial_s A(s)}.  
\end{align}

Equating the right-hand sides of Eqs. \ref{AR2} and \ref{AR3} then yields
\begin{align}\label{AR4}
& \arg\braket{\breve A(s_i) | \breve A(s_f)} -\imath \int_{s_i}^{s_f} ds \, \braket{\breve A(s) | \partial_s\breve A(s)}  \nonumber \\
&=\arg\braket{ A(s_i) | A(s_f)} -\imath \int_{s_i}^{s_f} ds \, \braket{A(s) | \partial_sA(s)} .
\end{align}

Define the dynamic phase as\cite{Aharonov_1987}
\begin{equation}\label{phidyn_single}
\Phi_{\rm dyn,A} = -\imath\int_{s_i}^{s_f} ds \braket{A(s) | \partial_{s}A(s)},
\end{equation}
so that Eq. \ref{AR4} can be written as
\begin{equation}\label{AR5}
\breve\Phi_{\rm tot,A} - \breve\Phi_{\rm dyn,A}  =\Phi_{\rm tot,A} - \Phi_{\rm dyn,A} .
\end{equation}

\begin{figure}[t]
	\begin{center}
		\includegraphics[width=0.5\linewidth]{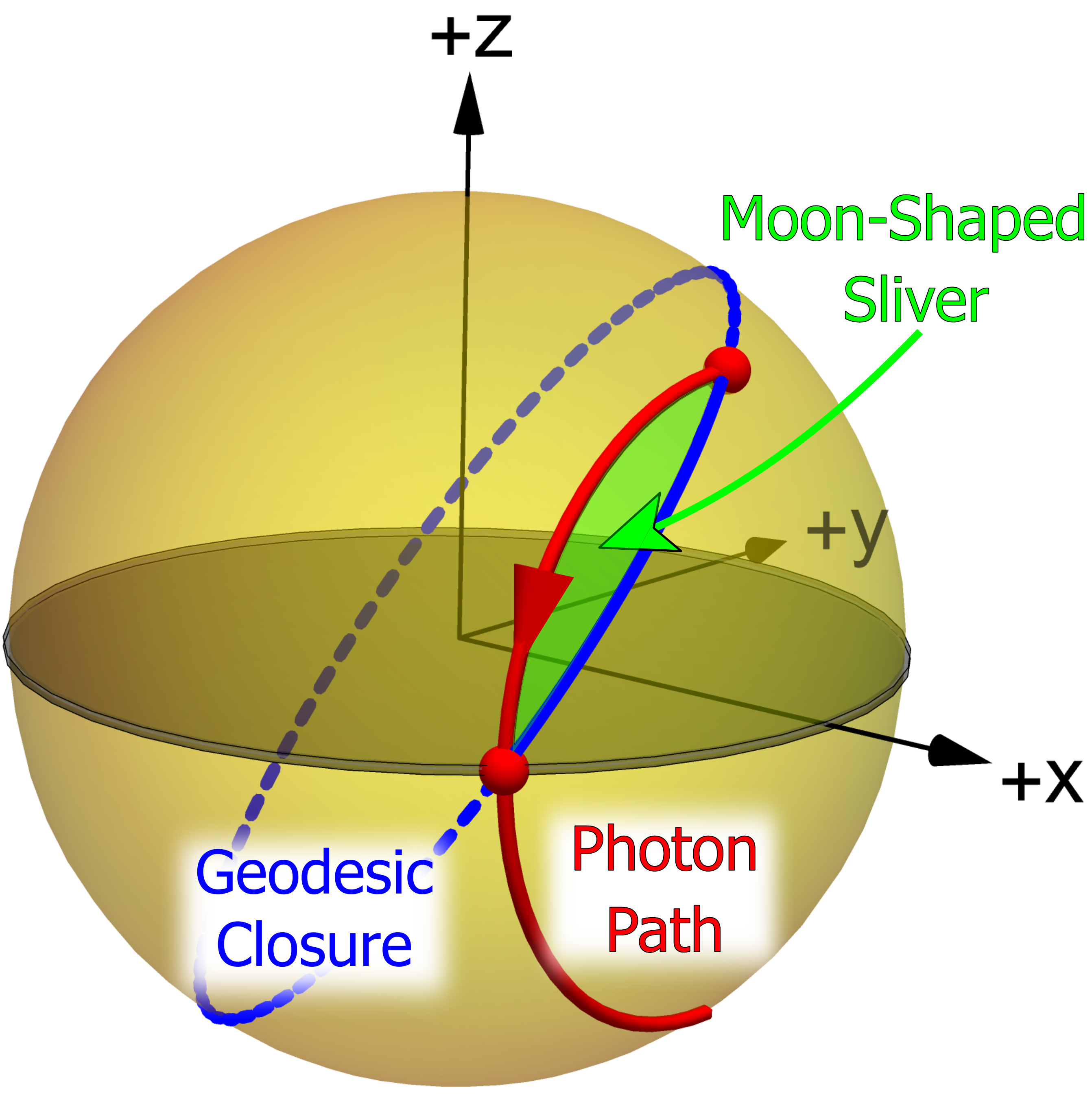}
	\end{center}
	\caption{ \emph{Single-Photon Geometric Phase on the SoM.} The geometric phase accumulated by single-photon transit through a lens is quantified using Eq. \ref{phigeom_single}. This has a geometric interpretation as the solid angle of a moon-shaped sliver, shown here, between the small-circle arc of the trajectory and a closure geodesic.} 
	\label{Solid_Angle}
\end{figure}

This implies that $\Phi_{\rm tot,A} - \Phi_{\rm dyn,A}$ is gauge-invariant, and we define this difference as the geometric phase:
\begin{equation}\label{phigeom_single}
\Phi_{\rm geom,A} = \Phi_{\rm tot,A} - \Phi_{\rm dyn,A} .
\end{equation}
Explicit dependence on $s$ has been suppressed here. Each point on a given trajectory can thus be associated with a gauge-invariant geometric phase\cite{Lusk_2022}, and a representative application is shown in Fig. \ref{Solid_Angle}. This formalism can be extended to pairs of entangled photons as well, to be taken up next.

%
%

\section{Entangled-Photon Holonomy}

\subsection{Construction of Two-photon States}
\begin{figure}[t]
	\begin{center}
		\includegraphics[width=0.5\linewidth]{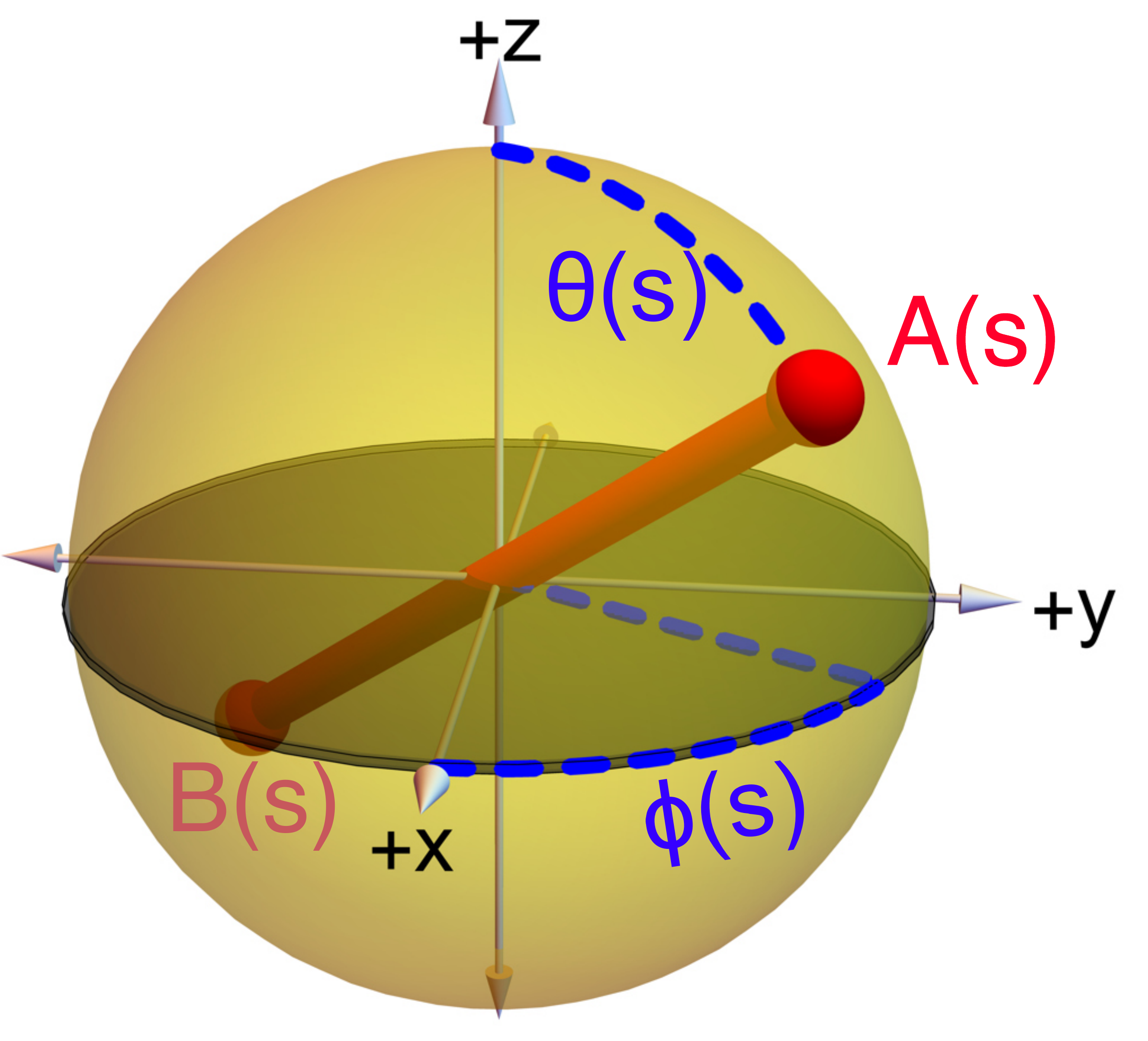}
	\end{center}
	\caption{ \emph{Components of Two-Photon States on the SoM.}  State $\ket{A(s)}$ and antipodal state $\ket{B(s)}$ are projected onto the first-order SoM for $\qi=0.6\pi/2$, $\phii=\pi/3$.} 
	\label{Tilted_States}
\end{figure}

The most general Schmidt form of an entangled two-photon state is\cite{Peres_1997}
\begin{equation}\label{Psi}
\ket{\Psi} = e^{-\imath\frac{\beta}{2}}\cos\frac{\alpha}{2}\ket{n} \ket{m} + e^{\imath\frac{\beta}{2}}\sin\frac{\alpha}{2}\ket{-n} \ket{-m},
\end{equation}
where $\ket{n}$ and $\ket{m}$ are arbitrary states on the SoM and $\ket{-n}$ and $\ket{-m}$ are their orthogonal counterparts. Here $\alpha$ characterizes the strength of entanglement while $\beta$ is the entanglement phase.  For the sake of simplicity in this initial investigation, though, we restrict attention to states in which $\ket{n}=\ket{m} =: \ket{A}$ and $\ket{-n} =: \ket{B}$. From Eq. \ref{stateA}, we then have the orthogonal (antipodal) state
\begin{equation}\label{stateB}
\ket{B} = e^{-\imath \chi} \left(e^{-\imath \frac{\phi}{2}} \sin\frac{\theta}{2}  \ketp - e^{\imath \frac{\phi}{2}} \cos\frac{\theta}{2}  \ketm \right) .
\end{equation}
Note that the antipodal relationship implies that horizontal lifts will be of opposite sign. Such two-photon states are illustrated in Fig. \ref{Tilted_States}.

The path of each photon through an optical circuit is unaffected by entanglement, so evolution can be described by (suppressing the explicit dependence on $s$)
\begin{equation}\label{Psi}
\ket{\Psi} = e^{-\imath\frac{\beta}{2}}\cos\frac{\alpha}{2}\ket{A} \ket{A} + e^{\imath\frac{\beta}{2}}\sin\frac{\alpha}{2}\ket{B} \ket{B} .
\end{equation}
At issue is the influence that entanglement has on holonomies associated with the evolution of such states. For notational convenience, label initial and final states as $\ket{A_i}$, $\ket{A_f}$, $\ket{B_i}$, and $\ket{B_f}$. The corresponding two-photon states are
\begin{align}\label{Psi_if}
\ket{\Psi_i} = &e^{-\imath\frac{\beta}{2}}\cos\frac{\alpha}{2}\ket{A_i} \ket{A_i}+ e^{\imath\frac{\beta}{2}}\sin\frac{\alpha}{2}\ket{B_i} \ket{B_i} \nonumber \\
\ket{\Psi_f} = &e^{-\imath\frac{\beta}{2}}\cos\frac{\alpha}{2}\ket{A_f} \ket{A_f}+ e^{\imath\frac{\beta}{2}}\sin\frac{\alpha}{2}\ket{B_f} \ket{B_f} .
\end{align}
%

%
%

\subsection{Entangled-State Projection, $\PPsi$}

The projection of initial state onto final state, $\PPsi := \braket{\Psi_i | \Psi_f}$, can be evaluated by expanding the inner product and expressing the result in terms of the following mode projections:
\begin{align}\label{projsAB}
\Paa&:=\braket{A_i | A_f}, \quad \Pbb:=\braket{B_i | B_f},\nonumber \\
\Pab&:=\braket{A_i | B_f}, \quad \Pba:=\braket{B_i | A_f}.
\end{align}
The form of Eqs. \ref{stateA} and \ref{stateB} immediately imply that
\begin{equation}\label{proj_rels}
\Pbb = \Paa^*, \quad \Pba = -\Pab^*.
\end{equation}
Keeping in mind that $\braket{B_i | A_i} = 0$ and $\braket{B_f | A_f} = 0$, we find that
\begin{align}\label{PE}
\PPsi = & \frac{\sin\alpha}{2}\left( e^{\imath \beta}\Pab^2 + e^{-\imath \beta} \Pba^2 \right)   \nonumber \\ 
&+ \cos^2\frac{\alpha}{2} \Paa^2 + \sin^2\frac{\alpha}{2} \Pbb^2 .
\end{align}
This result is made more meaningful by expressing it in terms of eigenvalues of the reduced state operator. Towards that end, define the state operator, $\hat\rho = \ket{\Psi}\bra{\Psi}$, then take the partial trace over either photon state to obtain
\begin{equation}\label{rhored}
\hat\rho_{\rm red} = \cos^2\frac{\alpha}{2}\ket{A}\bra{A} + \sin^2\frac{\alpha}{2}\ket{B}\bra{B} .
\end{equation}
It is clear that the eigenvalues of this reduced state operator are
\begin{equation}\label{evals}
\lambda_A = \cos^2\frac{\alpha}{2}, \quad \lambda_B = \sin^2\frac{\alpha}{2} .
\end{equation}
Therefore
\begin{equation}\label{PE2}
\PPsi =\frac{\sin\alpha}{2}\bigl(e^{\imath \beta} \Pab^2  + e^{-\imath \beta} \Pba^2 \bigr) + \lambda_A \Paa^2 + \lambda_B \Pbb^2 ,
\end{equation}
now in terms of standard measures of entanglement.

\subsection{Entangled-State Geometric Phase}
The entangled-state counterpart to the single-photon total phase of Eq. \ref{phitot_single} is
\begin{equation}\label{phitot_E}
\ptE = \arg(\PPsi).
\end{equation}
This is gauge-dependent as well, and the term that renders it invariant can be derived in exactly the same way as for Eq. \ref{AR5}. The entangled-state dynamic phase is therefore
\begin{equation}\label{phidyn_E}
\pdE= -\imath\int_{s_i}^{s_f} ds \braket{\Psi(s) | \partial_{s}\Psi(s)},
\end{equation}
where the evolving two-photon state, $\ket{\Psi(s)}$, is given in Eq. \ref{Psi}. Keeping in mind that the modes are orthogonal and that the projection and summation operations commute, we find that
\begin{equation}\label{phidyn_E2}
\pdE(s) = 2\lambda_A \Phi_{\rm dyn,A} + 2\lambda_B \Phi_{\rm dyn,B}.
\end{equation}

It is now possible to construct a geometric phase associated with two-photon states:
\begin{equation}\label{phigeom_E}
\pgE = \ptE - \pdE.
\end{equation}

As noted previously\cite{Sjoqvist_2000, Hessmo_2000}, Eqs. \ref{PE2}, \ref{phitot_E}, \ref{phidyn_E2}, and \ref{phigeom_E} imply that the geometric phase is a function of entanglement parameters $\alpha$ and $\beta$. However, these parameters pervade the equation structure making it difficult to elucidate the physics of this dependence. This motivates the consideration of ways to mine geometric phase in search of insight.

\section{Geometric Phase of Entanglement}

Since our interest lies in the influence of entanglement on geometric phase, introduce a \emph{Geometric Phase of Entanglement} ($\GPhi$) as
\begin{equation}\label{GPE1}
\GPhi := \pgE - (2 \lambda_A \pgA + 2 \lambda_B \pgB).
\end{equation}
$\GPhi$ is a linear combination of three gauge-invariant measures and so is itself gauge invariant. The term in parentheses is the mode-weighted geometric phase associated with single-photon transits, so $\GPhi$ represents a phase-based measure of the way in which entanglement influences holonomy. 

Each of the terms in Eq. \ref{GPE1} includes a path-dependent dynamic phase contribution, but Eq. \ref{phidyn_E2} shows that the dynamic phase of the entangled state is just the weighted sum of dynamic phases associated with the four modes that comprise the two-photon state. The factors of two are present because each of the four components accumulates a dynamic phase. This is the same result that we would obtain by carrying out a set of experiments in which a two-photon state is prepared in either $\ket{A_i}\ket{A_i}$ or $\ket{B_i}\ket{B_i}$, with the projection subsequently measured. Here the initial state is chosen using a random variable weighted with probabilities $2\lambda_A$ and $2\lambda_B$, respectively. The average of the measured dynamic phases is the same as that of the entangled state of Eq. \ref{phidyn_E2}. In other words, entanglement does not influence the accumulation of dynamic phase. Eq. \ref{GPE1} is therefore reduced to 
\begin{equation}\label{GPE2}
\GPhi = \arg\PPsi -2\bigl(\lambda_A \arg\Paa+ \lambda_B \arg\Pbb\bigr).
\end{equation}
This implies that the holonomic influence of entanglement depends only on the end states and not on the path taken, by itself an important conclusion.

$\GPhi$ exhibits two basic ways in which parameters $\alpha$ and $\beta$ affect holonomy. While these are used to construct entangled states, their influence on holonomy can be identified as either a \emph{superposition effect} or a \emph{quantum correlation effect}.  


The superposition effect is best demonstrated by considering a linear combination of separable product states. Define $s$-parametrized product states
\begin{equation}\label{AABB}
\ket{\mathbb A}:= \ket{A}\ket{A}, \quad \ket{\mathbb B}:= \ket{B}\ket{B}
\end{equation}
and their individual projections
\begin{equation}\label{PAABB}
\braket{\mathbb A_f | \mathbb A_i} \equiv \Paa^2, \quad \braket{\mathbb B_f | \mathbb B_i} \equiv \Pbb^2 .
\end{equation}
These find meaning via a set of gedanken experiments in which a two-photon state is prepared in either $\ket{\mathbb A_i}$ or $\ket{\mathbb B_i}$, with the projection subsequently measured. The initial state is chosen using a coin flip weighted with probabilities $\lambda_A$ and $\lambda_B$, respectively. The average of the measured projections defines the separable state projection,
\begin{equation}\label{PS}
\PS := \lambda_A \Paa^2 + \lambda_B \Pbb^2.
\end{equation}
While the eigenvalues are functions of entanglement parameter $\alpha$, as shown in Eq. \ref{evals}, this separable state projection, $\PS$, does not exhibit quantum correlation but, rather, a simple superposition of separated states. 

Armed with this example, now return to the $\GPhi$ of Eq. \ref{GPE2}. The second term is clearly a superposition effect. On the other hand, Eq. \ref{PE2} shows that the projection of initial and final entangled states, $\PPsi$, has elements of superposition (second and third terms) but also quantum correlation (first term).  All of the geometric quantum correlation exists within $\PPsi$, but that term is embedded within a nonlinear function which is only part of the overall construction. Focusing on $\GPhi$ instead of the geometric phase has allowed us to remove dynamic phase contributions, a substantial step in the right direction. However, the $\GPhi$ still obfuscates the role of entanglement on holonomy, and this motivates the construction of a different measure, one that both pinpoints and explains the geometric nature of quantum correlation.

\section{Geometric Projection of Entanglement}

The geometric manifestation of quantum correlation can be more clearly elucidated with a measure based on projection differences instead of a phase differences. Using Eqs. \ref{PE2} and \ref{PS}, the requisite projective difference is referred to as the Geometric Projection of Entanglement, $\Gproj$:
\begin{equation}\label{Gproj1}
\Gproj:= \PPsi - \PS = \frac{\sin\alpha}{2} \bigl(e^{\imath \beta} \Pab^2  + e^{-\imath \beta}\Pba^2 \bigr) .
\end{equation}
We immediately see that this measure does not include any superposition effects. As shown next, it is also gauge invariant.

Introduce the gauge transformation of a generic two-photon state,
\begin{equation}\label{Psixform}
\ket{\Psigengauge(s)} := e^{\imath \gammagen(s)}\ket{\Psigen(s)},
\end{equation}
with $\gammagen$ a smooth but otherwise arbitrary scalar function. Take the inner product of this field with itself at two positions to obtain
\begin{equation}\label{PsiR2}
\Projgengauge = e^{\int_{s_i}^{s_f} ds \, \partial_s \gammagen(s) )} \Projgen.  
\end{equation}
The subscript $0$ on all terms serves as a reminder that this is a generic two-photon state. Holding onto this relation for a moment, take derivative of Eq. \ref{Psixform} and then the inner product with $\Projgengauge$ to obtain
\begin{align}\label{PsiR3}
\int_{s_i}^{s_f} ds\partial_s \gammagen(s) = &-\imath \int_{s_i}^{s_f} ds \braket{\Psigengauge(s) | \partial_s\Psigengauge(s)} \nonumber\\
&+ \imath \int_{s_i}^{s_f} ds \braket{\Psigen(s) | \partial_s\Psigen(s)}.
\end{align}
The right-hand side is just the difference of two dynamic phases:
\begin{equation}\label{PsiR4}
\int_{s_i}^{s_f} ds\partial_s \gammagen(s) = \Phidyngengauge -\Phidyngen.
\end{equation}
Substitute this expression into Eq. \ref{PsiR2}, and re-arrange to obtain
\begin{equation}\label{PsiR5}
e^{\Phidyngengauge}\Projgengauge= e^{\Phidyngen} \Projgen.  
\end{equation}
This implies that $e^{\Phidyngen} \Projgen$ is a gauge-invariant representation of any two-photon projection.

Apply this to $\Pab$ by first calculating the dynamic phase of state $\ket{A(s)B(s)}$.
\begin{equation}\label{ABR1}
\PhidynAB = -\imath\int_{s_i}^{s_f} ds \braket{A(s)B(s) | \partial_{s}\left(A(s)B(s)\right)}.
\end{equation}
A straightforward calculation shows that this reduces to
\begin{equation}\label{ABR2}
\PhidynAB = \PhidynA + \PhidynB.
\end{equation}
However, the antipodal relationship between $\ket{A}$ and $\ket{B}$ implies that $\PhigeomA = -\PhigeomB$, and the form of the states given in Eqs. \ref{stateA} and \ref{stateB} imply that $\PhitotA = -\PhitotB$. Using Eq. \ref{AR5}, we therefore have that
\begin{equation}\label{ABR3}
\PhidynB = -\PhidynA.
\end{equation}
Together, Eqs. \ref{ABR2} and \ref{ABR3} we then have that $\PhidynAB = 0$ and $\PhidynBA = 0$. Applying these results to the gauge relation of Eq. \ref{PsiR5} gives the property that we sought to establish, that projections $\Pab$ and $\Pba$ are gauge-invariant and so $\Gproj$ as well. 

We have determined that the Geometric Projection of Entanglement, $\Gproj$ of Eq. \ref{Gproj1}, is a projective, gauge-invariant counterpart to the Geometric Phase of Entanglement, $\GPhi$ of Eq. \ref{GPE1}. By virtue of Eq. \ref{proj_rels},  $\Gproj$ is real-valued. It is path-independent, relying only on end states, and is therefore equal to zero if the initial and final positions on the SoM are the same. In the absence of entanglement, $\alpha=0$ or $\alpha=\pi$, $\Gproj = 0$.  Finally, and perhaps most informative, $\Gproj$ is extremized for $\theta_f = \theta_i$ and $\chi = 0$. 

$\Gproj$ quantifies the quantum correlation manifested within the two inner products of the initial and final states of antipodal modes. These would not be present if we simply summed the contributions of two separated photons or considered statistically weighted product states. It is only their quantum correlation that generates $\Gproj$.

\subsection{$\GPhi$ versus $\Gproj$ for Solo-Photon Transit}

Additional insight into our two geometric measures of entanglement are obtained by considering a revised dynamic for which photon $\#2$ does not transit a holonomy-producing circuit.  The entangled-state projection of Eqs. \ref{PE2} is modified to
\begin{align}\label{PE_solo1}
\PPsisolo = &\lambda_A \braket{A_f | A_i}\braket{A_i | A_i} + \lambda_B \braket{B_f | B_i}\braket{B_i | B_i} \nonumber \\ 
& + \frac{\sin\alpha}{2} e^{\imath \beta} \braket{A_f | B_i}\braket{A_i | B_i}  \\
& + \frac{\sin\alpha}{2} e^{-\imath \beta}\braket{B_f | A_i}\braket{B_i | A_i}. \nonumber
\end{align}
But $\braket{A_i | B_i} = \braket{B_i | A_i} = 0$, so this reduces to
\begin{equation}\label{PE_solo2}
\PPsisolo = \lambda_A \Paa + \lambda_B \Pbb.
\end{equation}
To calculate the associated $\Gproj$, an appropriate projection must also be identified to replace $\PS$ of Eq. \ref{PS}. Since photon \#1 generates projections $\Paa$ and $\Pbb$ while photon $\#2$ always has unit projection, the appropriate product state projection is
\begin{equation}\label{PS_solo}
\PSsolo = \lambda_A \Paa + \lambda_B \Pbb.
\end{equation}
$\Gproj$ is the difference between $\PPsisolo$ and $\PSsolo$, implying that 
\begin{equation}\label{GME_solo}
\Gprojsolo = 0.
\end{equation}
Both photons must traverse the circuit to produce a projection attributable to entanglement. 

The absence of $\Gproj$ is in sharp contrast to the finite $\GPhi$ produced by a general solo-photon transit.  Eq. \ref{GPE1} is modified to
\begin{align}\label{GPEsolo}
\GPhisolo = &\arg(\lambda_A \Paa + \lambda_B \Pbb)  \nonumber \\
&-\bigl(\lambda_A \arg\Paa+ \lambda_B \arg\Pbb\bigr),
\end{align}
where Eq. \ref{PE_solo2} has been used. This has been interpreted elsewhere \cite{Sjoqvist_2000, Hessmo_2000} as being a direct result of entanglement. While that is true, strictly speaking, the parametric dependence on $\alpha$ is only from $\lambda_A$ and $\lambda_B$. The same result could be achieved by weighting the projections of two separated photons that sequentially transit the same circuit. It is not really a manifestation of the interplay of quantum correlation and geometry but, rather, a consequence of mode superposition associated with an entangled state. The term is finite only because the operations of argument and summation do not commute. This setting of single-photon transits makes it clear why $\Gproj$ is more useful for understanding how quantum correlation influences holonomy. Attention can now be turned to the realization of these measures for a specific optical circuit.

\section{Generation of States with Tunable Entanglement}

The generation of $\GPhi$ and $\Gproj$ for an optical circuit calls for an initial two-photon states with tunable entanglement, Eq. \ref{Psi}. These can be engineered by constructing pump photons comprised of a linear combination of LG modes. We have devised a method for doing this, both explained and applied below.

\subsection{General Two-photon Entangled States}

The initial two-photon state of Eq. \ref{Psi_if} can be can be expanded and then expressed in the LG basis as
\begin{align}\label{Psi_w}
\Psi_i = &\quad w_{++}\ket{++} + w_{+-}\ket{+-} \nonumber \\
&+ w_{-+}\ket{-+} + w_{--}\ket{--},
\end{align}
where
\begin{align}\label{wfuncs}
w_{++} = 
& e^{-\imath \frac{\beta}{2}} \cos\frac{\alpha}{2} \cos^2\frac{\qi}{2} + e^{\imath \frac{\beta}{2}}\sin\frac{\alpha}{2}  \sin^2\frac{\qi}{2} \nonumber \\
w_{+-} = 
& \frac{1}{2}\sin\qi \left( -e^{\imath \frac{\beta}{2}} \sin\frac{\alpha}{2}  + e^{-\imath \frac{\beta}{2}} \cos\frac{\alpha}{2} \right) \nonumber \\
w_{-+} = 
& \frac{1}{2}\sin\qi \left( e^{-\imath \frac{\beta}{2}} \cos\frac{\alpha}{2}   - e^{\imath \frac{\beta}{2}} \sin\frac{\alpha}{2} \right) \\
w_{--} = 
& e^{\imath \frac{\beta}{2}} \sin\frac{\alpha}{2}  \cos^2\frac{\qi}{2} + e^{-\imath -\frac{\beta}{2}}\cos\frac{\alpha}{2} \sin^2\frac{\qi}{2}. \nonumber 
\end{align}
Here the initial azimuth angle is set to zero to more clearly focus on the essential physics.  Two-photon initial states on the first-order SoM, with arbitrary positions described by $\{\qi, 0\}$, therefore require four components of the OAM spectrum: $\{\el_1, \el_2\}= \{+1,+1\}, \{+1,-1\}, \{-1,+1\}, \{-1,-1\}$. The first and last of these are not present in the spectrum associated with the Type-I SPDC\cite{Chiao_1995} of a Gaussian beam. Because that is our chosen means of generating entangled states, the pump photon must be structured so that all four of these components can be produced within the birefringent crystal. 

\subsection{Pump Engineering with OAM Modes}

A quantum theory for SPDC has been developed \cite{Walborn_2010}, in which a classical Hamiltonian for the nonlinear interaction is subsequently quantized. The resulting Hamiltonian operator can then be used to approximate a time-evolution operator using the first two terms in a Dyson series. Application of the time-evolution operator to the initial pump state delivers an approximation of the entangled two-photon state. A set of simplifying assumptions then results in a tractable expression in which the pump beam is treated classically, $\Phi_{\rm pump}$, and the two-photon state emerging from a thin, properly oriented crystal is 
\begin{equation} \label{superposition_all}
\Psi= \sum_{\el_1, p_1}\sum_{\el_2, p_2} \mathbb{C}^{p_1,p_2}_{\el_1,\el_2} \ket{\el_1,p_1; \el_2,p_2}.
\end{equation}
As a reminder, $\{\el,p\}$ refers to the orbital and radial parameters of LG modes, while subscripts 1 and 2 identify the photons produced. The coefficients, $\mathbb{C}^{p_1,p_2}_{\el_1,\el_2}$, are evaluated by integrating over the radial cross-section, $\bf r_{\perp}$, of the beam at its waist using spatial representations of the LG modes, $LG^p_\el$: 
\begin{equation}\label{C_coeffs1}
\mathbb{C}^{p_1,p_2}_{\el_1,\el_2} = \int {\bf dr} _\perp \Phi_{\rm pump} ({\bf r}_\perp) LG^{p_1}_{\el_1}({\bf r}_\perp)^*LG^{p_2}_{\el_2}({\bf r}_\perp)^*.
\end{equation}
In the present work, attention is restricted to the construction of pump photons that can be described with a linear combination of non-radial LG modes---i.e. modes for which $p=0$:
\begin{equation}\label{pump1}
\Phi_{\rm pump}({\bf r}_\perp)  = \sum_{\el} {\mathbb D}_{\el} LG_{\el}({\bf r}_\perp).
\end{equation}
Here ${\mathbb D}_{\el}$ are coefficients that must be determined so as to produce the desired two-photon state of Eq. \ref{Psi_w}. Any two-photon states for which either radial parameters $p_1$ or $p_2$ are not both equal to zero can be disregarded, so the two-photon state of interest is described by coefficients
\begin{equation}\label{C_coeffs2}
C_{\el_1,\el_2} = \int {\bf dr} _\perp \sum_\el {\mathbb D}_{\el} LG_\el\bigr( LG_{\el_1} LG_{\el_2} \bigl)^* .
\end{equation}

We can also neglect modes generated by the crystal that do not have an OAM with magnitude equal to 1. Our goal is therefore to derive pump coefficients, ${\mathbb D}_{\el}$, such that
\begin{align}\label{C_coeffs3}
& \mathbb{C}_{1,1} = w_{++}, \mathbb{C}_{1,-1} = w_{+-} \nonumber \\
&\mathbb{C}_{-1,1} = w_{-+}, \mathbb{C}_{-1,-1} = w_{--} \, .
\end{align}

It is not clear, at the outset, that this is even possible or what constraints exist on such a solution. Certainly the $\mathbb C$-coefficients must satisfy internal relations at the very least. Towards this objective, Eq. \ref{C_coeffs2} can be simplified to
\begin{equation}\label{C_coeffs4}
\mathbb{C}_{\el_1,\el_2} = {\mathbb D}_{\el}\int {\bf dr}_\perp \sum_\el  LG_\el\bigr( LG_{\el_1} LG_{\el_2} \bigl)^*,
\end{equation}
and conservation of OAM implies that the integral at right is zero unless $\el_1 + \el_2 = \el$. When this condition is met, the azimuthal integral evaluates to $2\pi$. 

Next decompose the $LG_\el$ modes into a product of radial and azimuthal components,
\begin{equation}\label{LG_decomp1}
LG_\el( {\bf r}_\perp) = R_\el(r)\Phi_\el(\phi),
\end{equation}
where
\begin{equation}\label{LG_decomp2}
R_\el(r) = \frac{e^{-\frac{r^2}{2}}r^{|\el|}}{\sqrt{\pi |\el|!}}, \quad \Phi_\el(\phi) = e^{\imath \phi} .
\end{equation}
Noting that $R_\el$ is real-valued, we find that
\begin{equation}\label{D_coeffs1}
\mathbb{D}_{\el_1+\el_2} = \frac{\mathbb{C}_{\el_1,\el_2}}{f(\el_1,\el_2)},
\end{equation}
where
\begin{equation}\label{f}
f(\el_1,\el_2) := 2\pi \int dr\, r R_{\el_1+\el_2}(r) R_{\el_1}(r)R_{\el_2}(r).
\end{equation}
With $\mathbb C$-coefficients prescribed by Eq. \ref{C_coeffs3}, the $\mathbb D$-coefficients obtained from Eqs. \ref{D_coeffs1} and \ref{f} result in a structured pump photon that produces tunably-entangled states of the form given in Eqs. \ref{Psi_w} and \ref{wfuncs}.

As an aside, the quotient at right in Eq. \ref{D_coeffs1}, neither the numerator nor the denominator are simple functions of the sum, $\el_1 + \el_2$, but their quotient is. This amounts to a restriction on the $\mathbb C$-coefficients that can be produced. In particular, it implies that
\begin{equation}\label{C_coeffs5}
\arg(\mathbb{C}_{\el_1,\el_2}) = \arg(\mathbb{D}_{\el_1+\el_2}) .
\end{equation}
The form of the integral used to construct the  $\mathbb C$-coefficients and $f$ also implies that
\begin{equation}\label{C_coeffs6}
\mathbb{C}_{\el_1,\el_2} =\mathbb{C}_{\el_2,\el_1} .
\end{equation}
Eqs. \ref{C_coeffs5} and \ref{C_coeffs6} restrict the two-photon spectra that can be produced using this technique. These constraints do not impact the application of this method to generate the two-photon spectrum of  Eq. \ref{Psi_w} though. 

The following engineered pump structure is found to produce what we need:
\begin{equation}\label{pump2}
\Phi_{\rm pump} = {\mathbb D}_{-2} LG_{-2} + {\mathbb D}_{0} LG_{0} + {\mathbb D}_{+2} LG_{+2}.
\end{equation}
The requiste $\mathbb D$-coefficients are
\begin{align}\label{D_coeffs2}
{\mathbb D}_{-2} = &\frac{27}{8} \sqrt{\frac{\pi }{2}} e^{\frac{i
   \beta }{2}} \sin \left(\frac{\alpha
   }{2}\right) \cos^2\left(\frac{\theta_i}{2}\right) \nonumber \\
   &+\frac{27}{8} \sqrt{\frac{\pi
   }{2}} e^{-\frac{i \beta }{2}} \cos
   \left(\frac{\alpha }{2}\right) \sin
   ^2\left(\frac{\theta_i}{2}\right)\nonumber \\
{\mathbb D}_{0} = & \frac{9}{8} \sqrt{\pi } \sin \left(\theta_i\right) \left(e^{-\frac{i \beta }{2}} \cos
   \left(\frac{\alpha }{2}\right)-e^{\frac{i
   \beta }{2}} \sin \left(\frac{\alpha
   }{2}\right)\right) \nonumber\\
{\mathbb D}_{+2} = &\frac{27}{8} \sqrt{\frac{\pi }{2}} e^{-\frac{i
   \beta }{2}} \cos \left(\frac{\alpha
   }{2}\right) \cos ^2\left(\frac{\theta_i}{2}\right) \nonumber \\
   &+\frac{27}{8} \sqrt{\frac{\pi
   }{2}} e^{\frac{i \beta }{2}} \sin
   \left(\frac{\alpha }{2}\right) \sin
   ^2\left(\frac{\theta_i}{2}\right).
\end{align}

\begin{figure}[t]
	\begin{center}
		\includegraphics[width=1.0\linewidth]{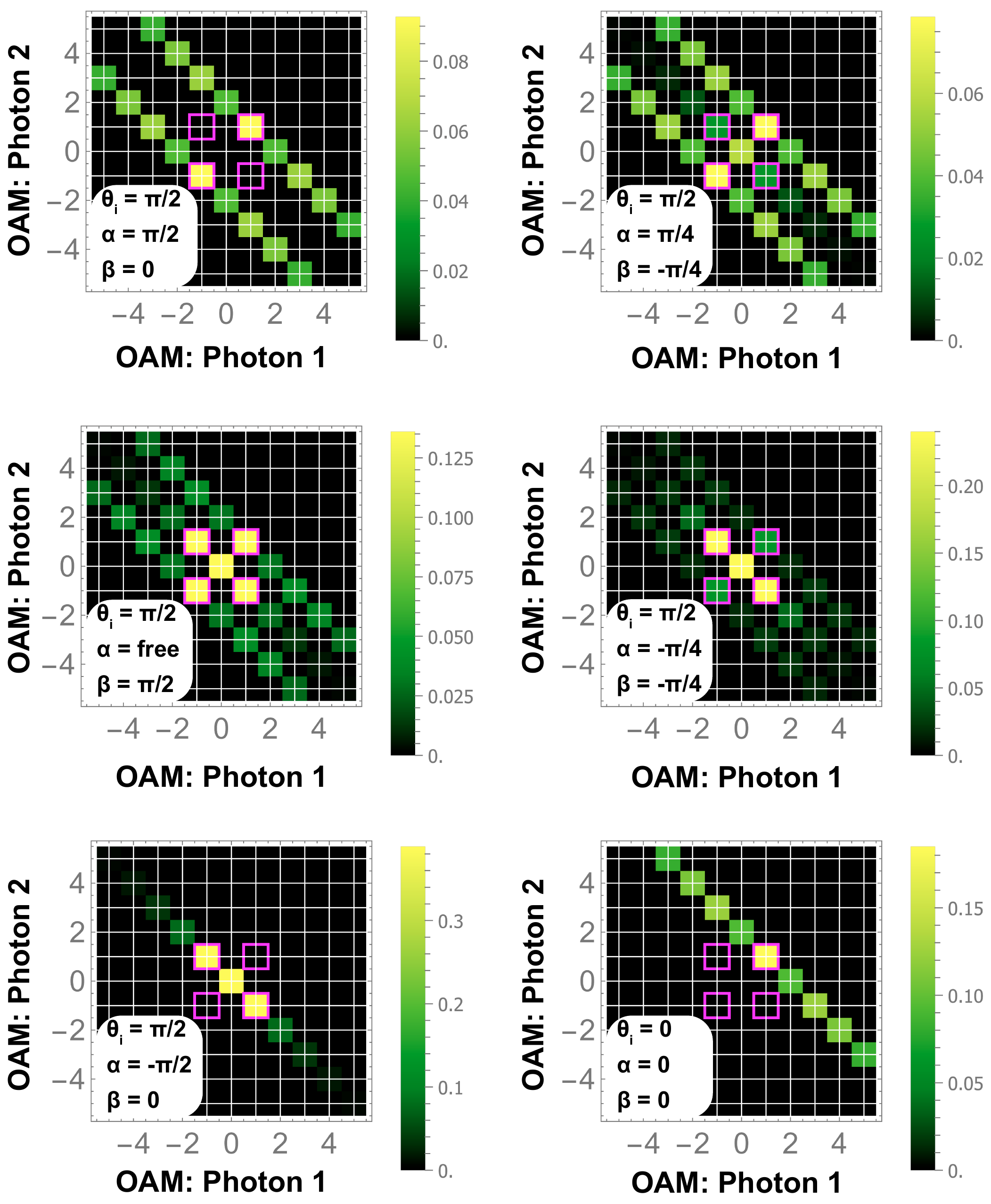}
	\end{center}
	\caption{ \emph{OAM Spectra Generated by Pump Engineering.}  Two-photon OAM magnitude spectra of states produced using pump engineering. The magenta frames identify first-order components. In all cases, initial azimuth angle $\phii=0$. The square of each magnitude gives the probability of measuring the associated state, normalized over all possible states for which radial LG parameter $p=0$.} 
	\label{OAM_Spectrum_++--}
\end{figure}

Representative spectra, generated with pump photons described by Eq. \ref{pump2}, are shown in Fig. \ref{OAM_Spectrum_++--}.

%
%

\section{Geometric Measures of Entanglement for a Pair of Mode Converters}

\subsection{Holonomy-Generating Optical Circuit}

\begin{figure}[t]
	\begin{center}
		\includegraphics[width=0.9\linewidth]{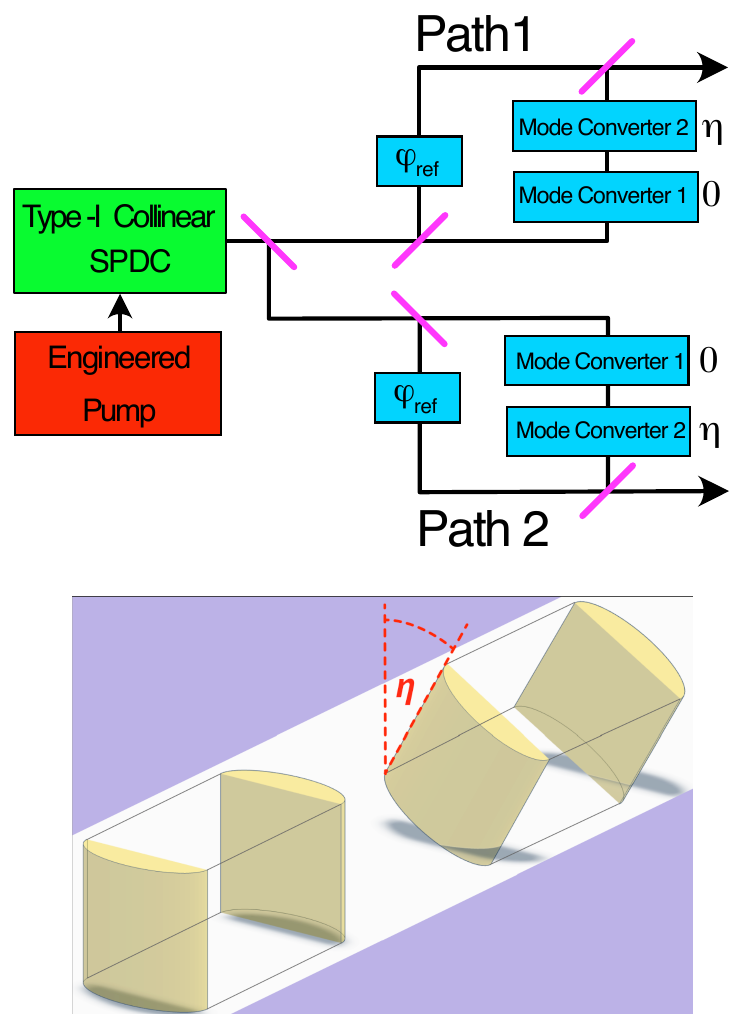}
	\end{center}
	\caption{ \emph{Schematic of Holonomy-Producing Optical Circuit.}  Upper Panel: Pump photons with an engineered spatial structure are used to produce entangled two-photon states, Eq. \ref{Psi}, with modes described by Eqs. \ref{stateA} and \ref{stateB}. Each photon is sent through a pair of $\pi$-mode converters\cite{Allen_1992, Beijersbergen_1993} for which the second converter is misoriented by angle $\eta$ relative to the first.  These paths are taken to be the same in the present work. Shifters used to remove phases accumulated in the beam splitters are not shown. Lower Panel: A pair of misoriented $\pi$-mode converters is shown to illustrate the misorientation angle, $\eta$. Each converter consists of a pair of astigmatic lenses.} 
	\label{Experimental_Schematic_2}
\end{figure}

We are now in a position to produce tunably entangled photons via pump engineering and then send the resulting bi-photon through a holonomy-generating optical circuit. This will allow the new geometric entanglement measures, $\GPhi$ and $\Gproj$, to be quantified and interpreted in a concrete setting.  A circuit sufficient for our purposes is shown schematically in Fig. \ref{Experimental_Schematic_2}.  Although the paths for each photon could be different, they are taken to be the same in the present work. 

\begin{figure}[t]
	\begin{center}
		\includegraphics[width=\linewidth]{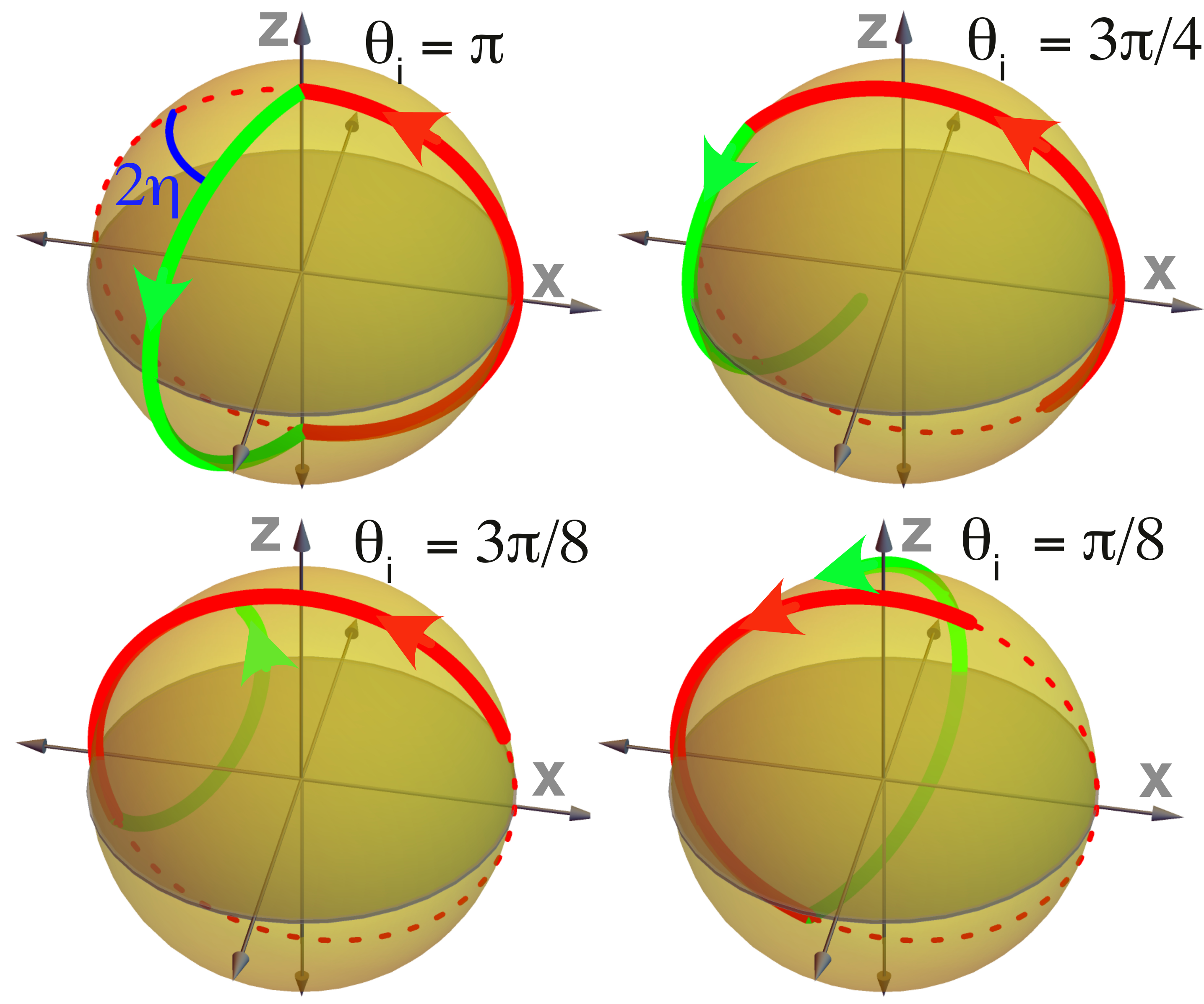}
	\end{center}
	\caption{ \emph{Photon Trajectories on the Sphere of Modes.} Trajectories composed of red and green geodesics are traced out by a photon in the $\ket{A}$ state with $\phii = 0$. The misorientation of the second $\pi$-converter is $\eta = \pi/6$ in all four cases shown. The angle between the continuation of the red curve (first mode converter) and the green curve (second mode converter) is $2\eta$, as shown in the upper-left panel.} 
	\label{Trajectories_on_SoM}
\end{figure}

Each photon is sent through two $\pi$-mode converters \cite{Allen_1992, Beijersbergen_1993}. These consists of a pair of astigmatic lenses, and we know that transit through the converters can be associated with trajectory paths on the SoM\cite{Voitiv_2022}. Representative trajectories are shown in Fig. \ref{Trajectories_on_SoM} for range of initial polar angles, $\qi$, all for the same angle of converter misorientation, $\eta = \pi/6$. 
%
%
%

Structured light can generate a phase holonomy as it propagates through the converters\cite{Voitiv_2022}. In fact, a meaningful holonomy can be associated even with lens transit fraction using a geometric notion of distant parallelism that does not require that the SoM trajectory be closed\cite{Lusk_2022}. Each photon subsequently passes through an spatial light modulator (SLM) and then a single-mode fiber (SMF) before being detected.  Higher-order and radial components of the superposition are disregarded because they will not contribute to the output of the SLMs, and the states are normalized accordingly. The single-mode fibers transmit only photons with a Gaussian structure, allowing the SLM/SMF combination to be viewed as selecting particular final states from the down-converted superposition\cite{Mair_2001}.

%
%
%

An analytical simulation is used to generate the evolving state associated with Fig. \ref{Experimental_Schematic_2}. From a modeling perspective, though, it is more straightforward to generate an equivalent analysis involving a sequence of four $\pi/2$-mode converters, implying that each photon travels through eight astigmatic lenses. For a given cylindrical lens, $j$, the input and output states are related by a linear operator, $\hat L^{{\rm\scriptscriptstyle Gouy}}_j$, that captures Gouy phase shifts\cite{Lusk_2022}:
\begin{equation}\label{Lens}
\ket{\psi^{(j)}_{\rm out}} = \hat L^{{\rm\scriptscriptstyle Gouy}}_j\ket{\psi^{(j)}_{\rm in}}.
\end{equation}
The converter misorientation, $\eta$, is accounted for using a rotation operator between lens $\#4$ and lens $\#5$. 

\subsection{$\GPhi$ for the Optical Circuit}

Two-photon state evolution is determined by evaluating Eq. \ref{Lens} to obtain the independent progression of single photons in modes $A$ and $B$ as they pass through the sequence of eight lenses. This delivers evolving states $\ket{A^{(j)}}(\zeta)$ and $\ket{B^{(j)}}(\zeta)$ for lens $j$, where $\zeta$ is the transit fraction through each lens, ranging from zero to one\cite{Lusk_2022}. The input state of lens $j+1$ is the output state of lens $j$. Intermediate states for lens $j$ then have the same structure as the two-photon state of Eq. \ref{Psi}:
\begin{align}\label{Psi_j}
\ket{\Psi_j(\zeta)} = &\quad e^{-\imath\frac{\beta}{2}}\cos\frac{\alpha}{2}\ket{A^{(j)}(\zeta)} \ket{A^{(j)}(\zeta)}  \nonumber \\
&+ e^{\imath\frac{\beta}{2}}\sin\frac{\alpha}{2}\ket{B^{(j)}(\zeta)} \ket{B^{(j)}(\zeta)}  .
\end{align}

The Geometric Phase of Entanglement, $\GPhi$, and the Geometric Projection of Entanglement, $\Gproj$, defined for general transits in Eqs. \ref{GPE2} and \ref{Gproj1}, can now be determined for the optical circuit of Fig. \ref{Experimental_Schematic_2}. Although these are functions of path parameter, $s$, here we focus only on their values after transit through the entire circuit. The projections that comprise these measures are found to be:
\begin{align}\label{Projs_circuit}
\Paa = & \braket{\scriptstyle A^{(8)}(1) | A^{(1)}(0)} = -\cos(2\eta) + \imath \sin(2\eta)\cos\qi \nonumber \\
\Pbb = &\braket{\scriptstyle B^{(8)}(1) | B^{(1)}(0)} = -\cos(2\eta) - \imath \sin(2\eta)\cos\qi \nonumber \\
\Pab = &\braket{\scriptstyle A^{(8)}(1) | B^{(1)}(0)} = -\imath \sin(2\eta)\sin\qi \nonumber \\
\Pba = &\braket{\scriptstyle B^{(8)}(1) | A^{(1)}(0)} = -\imath \sin(2\eta)\sin\qi .
\end{align}
Their application to Eqs. \ref{GPE2} results in the following rather opaque expression:
\begin{align}\label{GPE_circuit}
\GPhi &= \arg( T_R + \imath T_I) \nonumber \\
 & - \lambda_A\arg \left(\left(\cos (2 \eta )+i \sin (2 \eta ) \cos \left(\theta
   _i\right)\right){}^2\right) \nonumber \\
& - \lambda_A\arg \left(\left(\cos (2 \eta )-i \sin (2 \eta ) \cos \left(\theta
   _i\right)\right){}^2\right),
\end{align}
where
\begin{align}
T_R := &6 \cos (4\eta ) - 8 \sin\alpha \cos\beta \sin^2(2 \eta) \sin^2\left(\theta _i\right) + 2 \nonumber \\
&+ \cos \left(4 \eta - 2 \theta _i \right)+\cos\left(4 \eta +2 \theta _i\right)-2 \cos \left(2 \theta _i \right) \nonumber  \\
T_I := & 8 \cos\alpha  \sin (4 \eta ) \cos \left(\theta _i\right).
\end{align}
As expected, the Geometric Phase of Entanglement exhibits a complicated dependence on entanglement phase parameters, $\alpha$ and $\beta$, initial polar angle, $\theta_i$, and mode converter misorientation, $\eta$. Special cases can be visualized, but understanding the role of quantum correlation will have to wait until $\Gproj$ is considered.

As a prelude to visualizations of $\GPhi$, it is useful to quantify entanglement using Schmidt number, $K$, the effective number of modes in a Schmidt decomposition. In turn, $K$  can be defined in terms of traces associated with the reduced state operator and, ultimately, its eigenvalues from Eq. \ref{evals}:
\begin{equation}\label{K1}
K = \frac{\rm{Tr}^2\left(\hat{\rho}_{\rm{red}}\right)}{\rm{Tr}\left(\hat{\rho }_{\rm{red}}^2\right)} \equiv \frac{1}{\lambda_A^2 + \lambda_B^2} .
\end{equation}
For our optical circuit, this reduces to
\begin{equation}\label{K1}
K = \frac{4}{3+\cos(2\alpha)}.
\end{equation}
The relationship can be inverted piecewise to produce $\alpha$ as a function of $K$. Maximum entanglement is associated with $\alpha=\pi/2$ and $\alpha=3\pi/2$. For each of these, $K = 2$, implying that are two significant modes in the Schmidt decomposition. 

\begin{figure}[t]
	\begin{center}
		\includegraphics[width=0.7\linewidth]{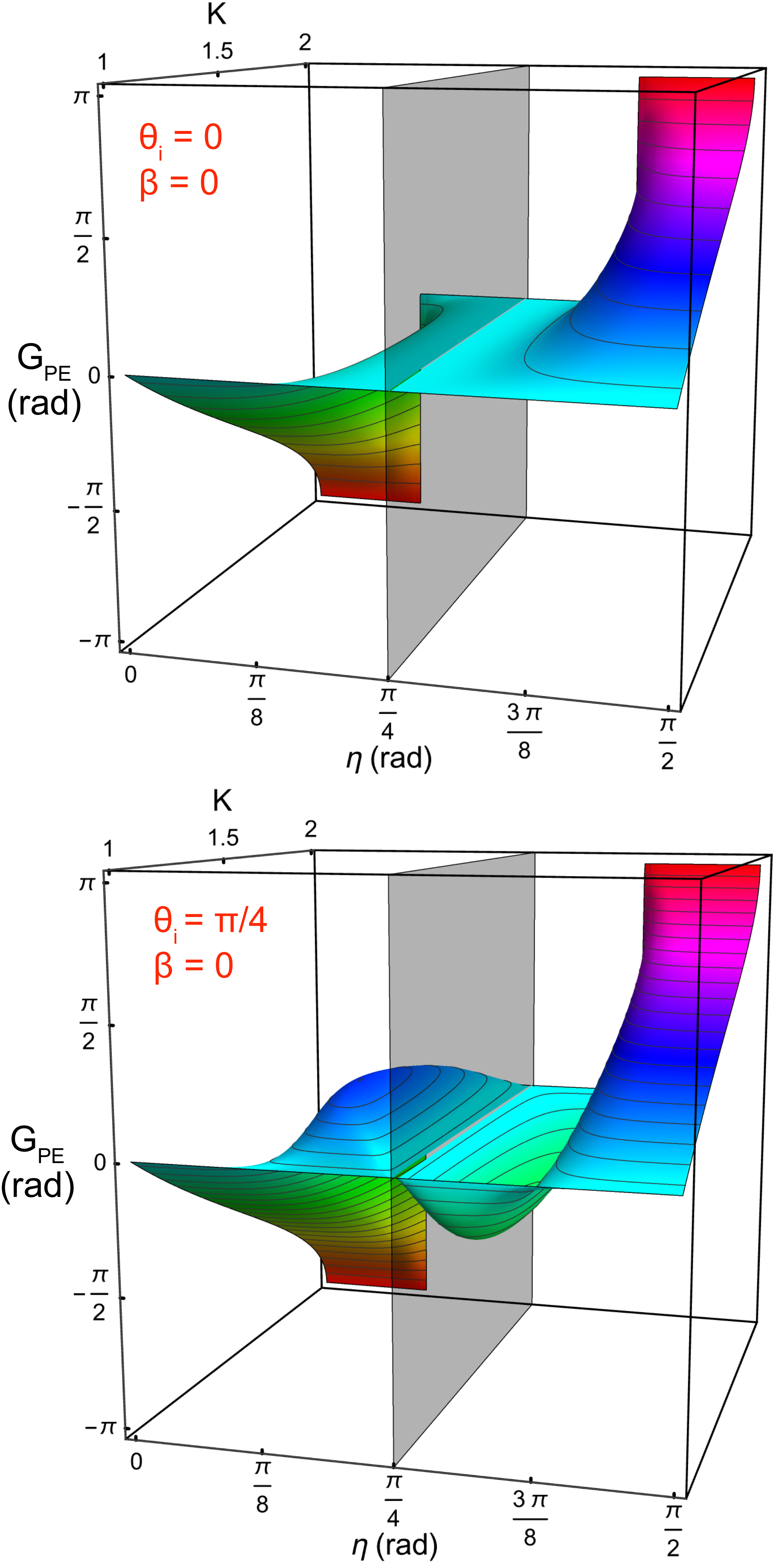}
	\end{center}
	\caption{ \emph{Geometric Phase of Entanglement for Two-Photon Transits Through Optical Circuit.} The $\GPhi$ of Eq. \ref{GPE_circuit} and Fig. \ref{Experimental_Schematic_2} is plotted as a function of mode converter misorientation angle, $\eta$, and Schmidt factor, $K$, with both initial polar angle, $\theta_i$, and entanglement phase parameter, $\beta$, prescribed and held fixed. Gray planes at $\eta=\pi/4$ are intended to help guide the eye.}
\label{GPE_Both_Photons_Transit_v4A}
\end{figure}

\begin{figure}[t]
	\begin{center}
		\includegraphics[width=0.7\linewidth]{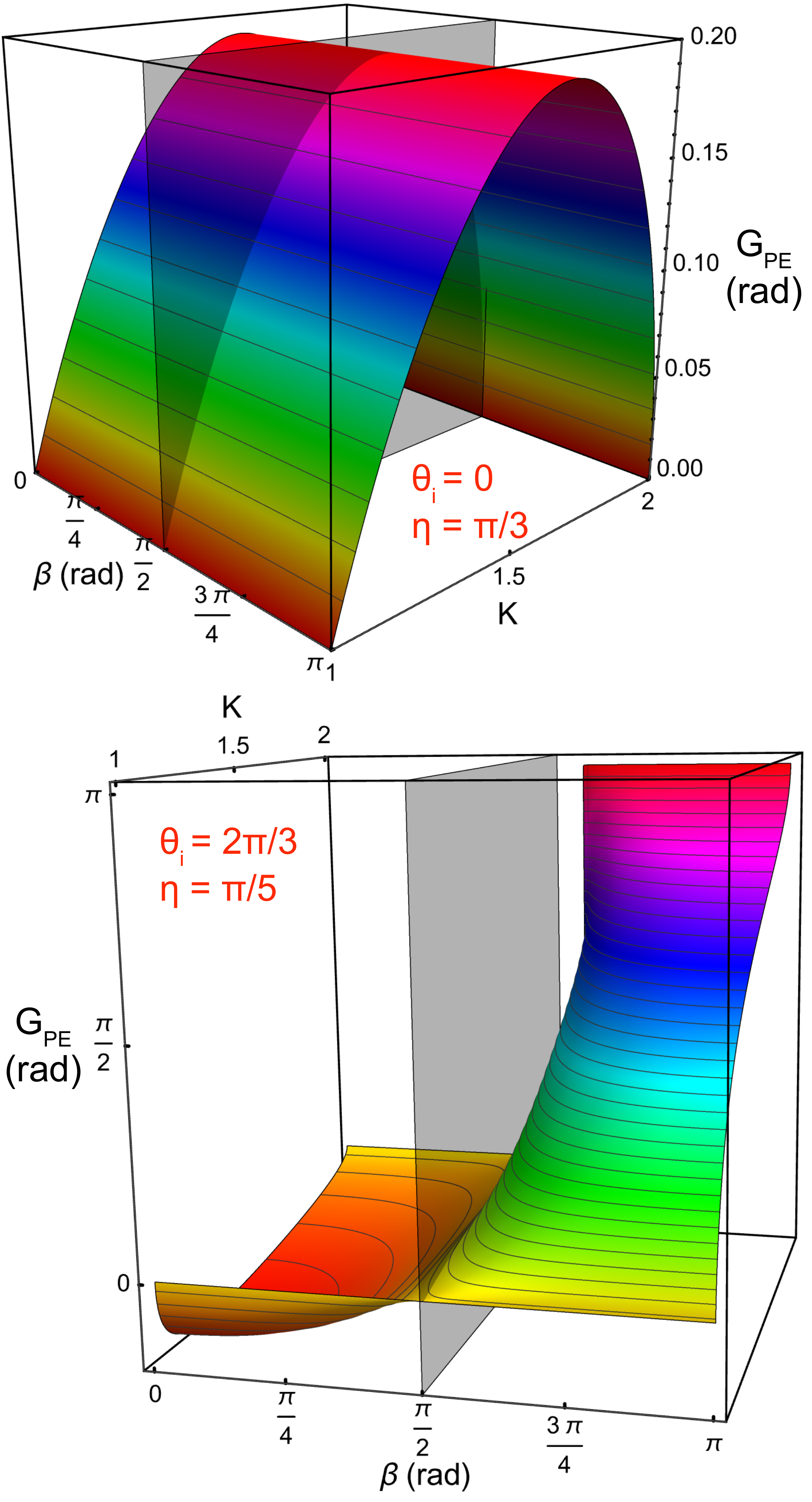}
	\end{center}
	\caption{ \emph{Geometric Phase of Entanglement for Two-Photon Transits Through Optical Circuit.} The $\GPhi$ of Eq. \ref{GPE_circuit} and Fig. \ref{Experimental_Schematic_2} is plotted as a function of entanglement phase parameter, $\beta$, and Schmidt factor, $K$, with both the initial polar angle, $\theta_i$, and mode converter misorientation angle, $\eta$, prescribed and held fixed. Gray planes at $\beta=\pi/2$ are intended to help guide the eye.}
\label{GPE_Both_Photons_Transit_v4B}
\end{figure}

The $\GPhi$ of Eq. \ref{GPE_circuit} is plotted in Figs. \ref{GPE_Both_Photons_Transit_v4A} and \ref{GPE_Both_Photons_Transit_v4B}  for four special cases. These collectively show that this geometric measure of holonomy is strongly influenced by the number of Schmidt modes. At the same time, they make it clear that $\GPhi$ is not a simple function of parameters $K$, $\beta$, $\eta$, or $\theta_i$. In all cases and as expected, $\GPhi=0$ when the second mode converter is not misoriented or if there is only a single mode---i.e. $K=0$. 

%
%
%

%
%
%

\subsection{$\Gproj$ for the Optical Circuit}

\begin{figure}[t]
	\begin{center}
		\includegraphics[width=0.8\linewidth]{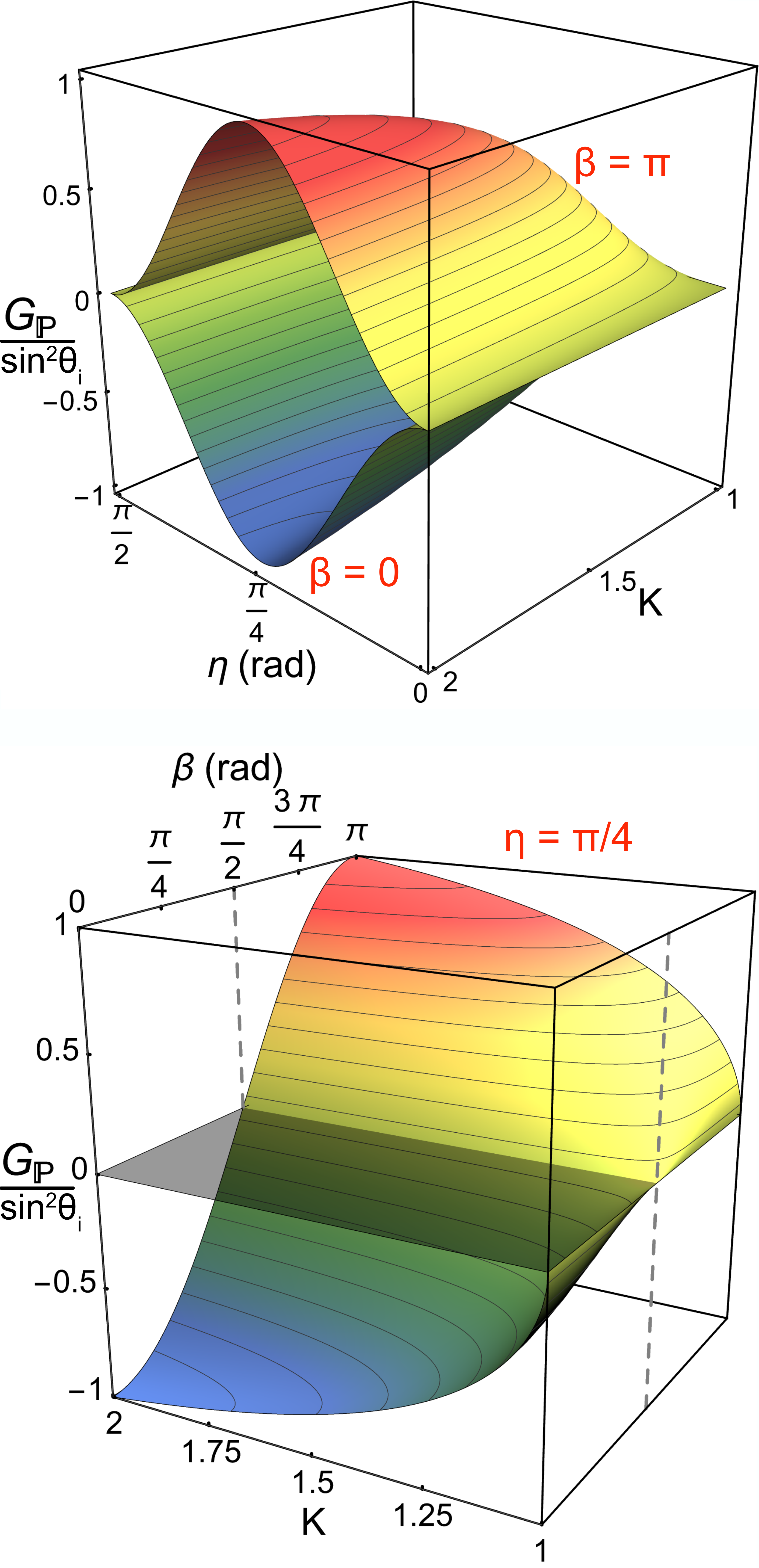}
	\end{center}
	\caption{ \emph{Geometric Projection of Etanglement.} The $\Gproj$ of Eq. \ref{Gproj1} and Fig. \ref{Experimental_Schematic_2} is plotted in two ways. In the top panel, $\beta=0$ and $\beta=\pi$, with $\Gproj$ is shown as a function of mode converter misorientation, $\eta$, and Schmidt number, $K$. In the bottom panel, the mode converter misorientation, $\eta=\pi/4$, and $\Gproj$ is shown as a function of entanglement phase, $\beta$, and Schmidt number, $K$.} 
	\label{GME_3D_Plots}
\end{figure}

In contrast to $\GPhi$, an application of Eqs. \ref{Projs_circuit} to Eq. \ref{Gproj1} results in a clean expression for the geometric manifestation of quantum correlation, $\Gproj$: 
\begin{equation}\label{GQC}
\Gproj = -\sin\alpha  \cos \beta \sin^2\qi \sin^2(2 \eta) .
\end{equation}
This is visualized with the pair of surface plots shown in Fig. \ref{GME_3D_Plots}. They elucidate the dependence of $\Gproj$ on  Schmidt number, $K$, initial polar angle, $\qi$, entanglement phase parameter, $\beta$, and mode converter misorientation angle, $\eta$. A quick observation confirms that $\Gproj = 0$ if any of the following is true: $\beta=\pi/2$, $\eta = 0$, or $\qi=0$. 
%
%
%
%

Several observations can now be made. Focus first on the role of the mode converter misalignment by holding the Schmidt number, $K$, entanglement phase, $\beta$, and initial polar angle on the SoM, $\qi$, all fixed. $\Gproj$ is then extremized by converter misorientations that cause the final state of mode $A(B)$ to have rotated into the initial state of mode $B(A)$. It also follows that $\Gproj=0$ when there is no misalignment, because then the final mode of $A(B)$ is orthogonal to the initial mode of $B(A)$. These relationships are made clear by the projection expressions of Eq. \ref{Projs_circuit}. Geometric quantum correlation arises from the projection of the initial state of one mode onto the final state of another. It is a signature feature of entanglement that these projections exist within the overlap of initial and final two-photon states.  The magnitude of $\Gproj$ is a maximum when the second mode converter has a misorientation of $\eta=\pm\pi/4$, because then the cross-term projections have a magnitude of one. If the initial states lie on the SoM poles, though, mode converter misorientation will not cause the desired rotation of the mode states, and the cross-projection terms of Eq. \ref{Projs_circuit} will be zero. $\Gproj$ is then zero as well. 

It also makes sense that the magnitude of $\Gproj$ is the largest when the modes are maximally entangled---i.e. when $K=2$. Likewise, $\Gproj=0$ when there is no entanglement $(K=1)$.

The lower plot in Fig. \ref{GME_3D_Plots} make it clear that $\Gproj$ is sensitive to the entanglement phase parameter, $\beta$. It is the smallest when $\beta = 0$, largest when $\beta = \pi$, and zero when $\beta=\pi/2$. This can be explained by referring back to Eq. \ref{Gproj1}. There we see that $\Gproj$ is comprised of the weighted sum of two non-negative projection terms, and Eq. \ref{Projs_circuit} shows that the terms themselves are the same for our optical circuit. Since they are weighted by $e^{+\imath \beta}$ and $e^{-\imath \beta}$, the two projection terms effectively cancel each other when $\beta=\pi/2$. We still have two contributions to $\Gproj$, but they are of opposite sign.

\section{Conclusions}

Structured light may exhibit a geometric phase holonomy when sent through an optical circuit. The same is true for structured bi-photons that are entangled, and quantum correlation has an influence on the associated holonomy. However, it is challenging to understand how this occurs because the entanglement parameters are also embedded in terms identifiable as statistical mixtures of separated photon modes. Some of the inscrutability can be removed by constructing a gauge-invariant Geometric Phase of Entanglement, $\GPhi$, which removes dynamic phase and path dependence from consideration. Even so, the role of quantum correlation remains abstruse. The remedy is to construct a gauge-invariant, real-valued, projection-based measure: the Geometric Projection of Entanglement, $\Gproj$.

$\Gproj$ offers clarity in pinpointing the way in which quantum correlation affects the manifestation of holonomy. After enumerating key general properties, it was subsequently evaluated for optical circuits comprised of two misoriented $\pi$-mode converters, allowing the measure to be visualized as a function of initial polar angle, converter misorientation angle, and Schmidt number. This was facilitated by a novel pump engineering method developed for the generation of photon pairs with tunable entanglement of first-order Laguerre-Gaussian modes. 

Within this particular setting, it was found that $\Gproj$ can be represented by a simple expression that is extremized when two pairs of mode converters are misoriented such that the final state of one mode overlays the initial state of the other. This is a signature feature of quantum correlation and lack of local-realism. It follows that $\Gproj=0$ when there is no misalignment, because then the final state of one mode is orthogonal to the initial state of the other. The magnitude of $\Gproj$ is the largest when the modes are maximally entangled. Likewise, $\Gproj=0$ when there is no entanglement. 

Since the Geometric Projection of Entanglement is gauge invariant, it has the stature of an experimentally measurable property of the end states of entangled-photon dynamics. Conveniently, $\Gproj$ is real-valued as well. The study of its accumulation as a function of path position in parameter space is a promising direction for future work. Our theoretical framework can also be extended to more general two-photon states and circuits, offering the prospect of developing a clearer understanding of the physical nature and potential quantum sensing applications of geometric quantum correlation. For instance, the entangled photons could traverse distinct SU(2) circuits, allowing for the study of an Aharonov-Bohm effect\cite{Aharonov_1959}. The consideration of radial Laguerre-Gaussian modes would also seem to offer a rich additional dimension for studies of the geometric, holonomic nature of entanglement.

\section{Acknowledgments}
It is a pleasure to acknowledge useful discussions with Andrew Voitiv and Mark Siemens.


\end{document}